# High-pressure cupric oxide: a room-temperature multiferroic

Xavier Rocquefelte[1*], Karlheinz Schwarz[2], Peter Blaha[2]

Sanjeev Kumar[3], Jeroen van den Brink[4]

[1]. *Institut des Matériaux Jean Rouxel, UMR 6502 CNRS – Université de Nantes, Boîte Postale 32229, 44322 NANTES Cedex 3, France*

[2]. *Institute of Materials Chemistry, Vienna University of Technology, Getreidemarkt 9/165-TC, A-1060 Vienna, Austria*

[3]. *Indian Institute of Science Education and Research Mohali, Knowledge City, Sector 81, Mohali 140306, India*

[4]. *Institute of Theoretical Solid State Physics, IFW Dresden, 01171 Dresden, Germany*

**Multiferroic materials, in which ferroelectric and magnetic ordering coexist, are of fundamental interest for the development of multi-state memory devices that allow for electrical writing and non-destructive magnetic read-out operation[1,2]. The great challenge is to create multiferroic materials that operate at room-temperature and have a large ferroelectric polarization P. Cupric oxide, CuO, is promising because of its large P ~ $10^2$ μC.m$^{-2}$, but is unfortunately only multiferroic in a temperature range of 20 K, from 210 to 230 K[3,4]. Here, using a combination of density functional theory and Monte Carlo calculations, we establish that pressure-driven phase competition induces a giant stabilization of**

**the multiferroic phase of CuO, which at 20-40 GPa becomes stable in a domain larger than 300 K, from 0 to T > 300 K. Thus, under high-pressure, CuO is predicted to be a room-temperature multiferroic with large polarization.**

Since the first observation of multiferroicity in CuO by Kimura and co-workers[4] it has been established that (i) CuO is a type-II multiferroic, so that ferroelectricity occurs as a result of magnetic ordering[3,4] and therefore the multiferroic ordering temperature equals the magnetic ordering temperature $T_N$ = 230 K; (ii) CuO is a quasi-one-dimensional magnetic system with a large magnetic coupling $J_z$ ~ 80 meV[5-7], which explains the high ordering temperature $T_N$; (iii) upon cooling, a polar incommensurate antiferromagnetic (AF) spin-spiral ordering, referred to as $AF_2$, appears below $T_N$ = 230K and a non-polar commensurate AF spin structure, $AF_1$, below the lock-in temperature $T_L$ = 213 K and (iv) the Dzyaloshinskii-Moriya (DM) "cycloidal" interactions play a major role in the emergence of the electric polarisation in CuO[8-10]. Different aspects of the interplay between the magnetic, orbital and electronic degrees of freedom in CuO have been studied intensely[8-14]. Recently we have shown that by applying a pressure of 8.8 GPa to CuO[14], the magnetic exchange interactions can increase by 46%. This holds the promise that under pressure $T_N$ will increase, perhaps even to room-temperature. Indeed, the monoclinic phase of CuO is known to be stable up to at least 70 GPa[15], even if detailed structural refinements are only available at pressures lower than 10 GPa[16]. Establishing the stability of the multiferroic phase under pressure, then, not only requires a calculation of the magnetic exchange interactions by density functional theory (DFT) but also a determination of $T_L$, $T_N$ and the temperature dependence of the polarisation P by complementary methodologies. For this we employ both a semi-empirical ansatz as well as unbiased classical Monte Carlo simulations.

CuO consists of corner- and edge-sharing square-planar $CuO_4$ units, which form $(\text{-Cu-O-})_\infty$ zigzag chains running along the [10-1] and [101] directions of the unit cell[17].

The low-T magnetic structure, $AF_1$, consists of Cu moments arranged antiferromagnetically along [10-1] and ferromagnetically along [101], with [010] direction as the easy axis[18]. The exchange interactions are captured by the Heisenberg Hamiltonian $H_H = \sum_{ij} J_{ij} \boldsymbol{S_i} \cdot \boldsymbol{S_j}$ which, to properly describe the magnetic properties of CuO, requires at least five magnetic exchange coupling parameters, *i.e.* four superexchange interactions ($J_a$, $J_b$, $J_x$, $J_z$) and one super-superexchange interaction ($J_{2a}$)[7,14,19,20], see Fig. 1a. The pressure dependence of the unit cell volume and of the $J$ values is shown in Figs. 1b and 1c, respectively.

The predictive power of our DFT geometry optimisation is confirmed by its capacity to reproduce the volume decrease with pressure as reported up to 17 GPa for nano-crystalline CuO samples[21]. The pressure dependence of the $J$ values, determined for the optimised atomic structures, compare very well to previous calculations[14], that use the available experimental structures up to 8.8 GPa[16]. Most importantly, $J_z$ strongly increases while $J_{2a}$ is nearly constant for pressures up to ~20 GPa, after which they increase in a similar manner as $J_z$. Of the three smaller $J$ values, $J_a$ is most affected by pressure and becomes ferromagnetic beyond about 2 GPa. The magnetic frustration, *i.e.* the competition between $J_a$ and $J_{2a}$[14], is therefore strongly enhanced by pressure. The change in ratio between the two largest $J$ values, *i.e.* $J_z/J_{2a}$, evidences that the effective magnetic dimensionality is also affected by pressure. As shown in Fig. 2a the quasi-1D character of the magnetic structure is enhanced for pressures up to 20 GPa and then reduced.

Having determined the pressure dependence of the magnetic exchange coupling constants, we calculate the multiferroic ordering temperature $T_N$ up to 200 GPa. We first evaluate it using the semi-empirical random phase approximation (RPA) expression for the quasi-1D antiferromagnetic Heisenberg model on a cubic lattice with intrachain and interchain couplings $J$ and $J'$, respectively[22] (see supplementary materials). The resulting $T_N$ is shown in Fig. 2b. Choosing the parameterization such that it reproduces the transition temperature $T_N$ = 230 K at ambient pressure, we observe a monotonic and substantial increase of $T_N$ with pressure. It coincides with the experimental pressure dependence of $T_N$, as was measured up to 1.8 GPa[23] and reaches room temperature at ~20 GPa. To substantiate this prediction, however, one needs to go beyond the semi-empirical approach. For this purpose we employed a classical Monte Carlo (MC) technique to explore the competition between the different magnetic states as a function of both pressure and temperature, with the Hamiltonian $H = H_H + H_{UA} + H_{DM} + H_{MA}$, where, $H_H$ is the Heisenberg exchange, $H_{UA}$ is uniaxial anisotropy (UA), $H_{DM}$ the Dzyaloshinskii Moriya (DM) term and $H_{MA}$ is the multiaxial (MA) anisotropy term. All these terms are relevant, but in particular the anisotropy terms are shown by our DFT calculations to be crucial to describe the effects of pressure. The relevant magneto-crystalline anisotropy energies (MAE) of CuO have been calculated for the ground-state $AF_1$ magnetic structure with MAE = E[uvw] – E[010], where E[uvw] is the energy deduced from spin-orbit calculations with magnetization along the [uvw] crystallographic direction. Fig. 3a shows the anisotropy energy surface[24] for CuO in the $AF_1$ magnetic order at a pressure of 0 GPa. Two minima are observed along [010] direction and equivalently [0-10] direction. Thus, the spin-orbit DFT calculations properly predict that the b-axis is the easy axis of magnetization of CuO for the low-T magnetic phase $AF_1$. A similar result is obtained for the entire pressure domain, *i.e.* from 0 to 200 GPa. However, the MAE values are rapidly decreasing with pressure as evidenced in Fig. 3b, in which the MAE in the (a,c)-plane is plotted as a function of the angle $\varphi$, such that $\varphi = 0°$ corresponds to the [101] direction. It also turns out that the hardest axis of magnetization (largest MAE value) is close to the [10-1] direction, *i.e.* the AFM direction. The pressure dependence of the MAE approximately follows an exponential decay, as is illustrated in Fig. 3c for the [-101] direction.

Before discussing the MC data, we can estimate the order of magnitude of the ferroelectric polarization, P, by using the empirical formula proposed by Katsura *et al.*[25]: $P = (V/\Delta)^3$, where V is the Cu-O electronic overlap integral and $\Delta$ is the p-d splitting. The superexchange parameter is approximately given by, $J = V^4/\Delta^3$. The relevant superexchange interactions for the ferroelectric nature of CuO are $J_a$ and $J_b$. Therefore taking J = 5 meV and $\Delta$ = 1.4 eV[7,26], we find P = 0.015 $\mu C.cm^{-2}$, which is very close to the experimental value. The Monte Carlo (MC) simulations use the *J* values obtained from the DFT calculations and, in particular, incorporate the rapidly decreasing $H_{MA}$ term with pressure. Fig. 4a shows the resulting spin current as a function of pressure, which is a quantity that is directly proportional to P. We observe that at ambient pressure close to the paramagnetic (PM) to AF1 transition, a spontaneous polarisation is induced. This polarisation is found to be non-zero between $T_N$ = 200 and $T_L$ = 150 K, which compares well to the experimentally observed stability domain of the incommensurate $AF_2$ magnetic order, between $T_N$ = 230 and $T_L$ = 213 K.

The fact that the calculated values are somewhat lower than the experimental ones is due to the model approximations involved and indicates that the MC results are conservative in the sense that they rather tend to underestimate the stability of the multiferroic phase. When the pressure is increased the polarisation grows and extends to a larger temperature range. At 30 GPa for instance an increase of about 20% is observed with respect to the polarisation at 0 GPa, and the temperature range is larger and in between 245 and 115 K. At 200 GPa, the multiferroic phase ($AF_2$) extends down to zero temperature and the ferroelectric polarisation is more than doubled. The MC results confirm the increase of $T_N$ with pressure, in accordance with the experimental observations for pressures up to 2 GPa and the results from the semi-empirical RPA expressions. We find good quantitative agreement for the values of the differential pressure increase of $T_N$ from experiment, 2.7 (0.2) K/GPa[23], and from the RPA and MC results, 3.5 (0.3) K/GPa and 3.0 (0.3) K/GPa, respectively. The calculated temperature-pressure phase diagram of CuO (see Fig. 4b) shows in addition a monotonic decrease of $T_L$ with pressure. As a consequence, the non-polar $AF_1$ phase disappears from the phase diagram with increasing pressure, at the benefit of the multiferroic $AF_2$ phase. The MC simulations indicate that $T_N$ reaches RT at ~40 GPa, which is higher than the ~20 GPa, obtained from the semi-empirical RPA expressions, underlining that the Monte Carlo pressure of ~40 GPa is a conservative estimate for the critical pressure value.

Finally, the present temperature-pressure phase diagram of CuO evidences a large increase of the stability range of the incommensurate multiferroic $AF_2$ phase under high-pressure, which is stable in a domain of only 20 K (from 210 to 230 K) at 0 GPa, and in a domain larger than 300 K (from 0 to T > 300 K) at 20-40 GPa. Such a giant stabilization of a multiferroic phase by pressure has never been observed and proposed. Indeed, except for CuO, all the reported pressure-temperature phase diagrams of multiferroic materials ($Ni_3V_2O_8$, $MnWO_4$, $TbMnO_3$, $RMn_2O_5$ with R = Tb, Dy, Ho…)[27–30] lead to the same conclusion: the stability range of the incommensurate magnetic phase is reduced by pressure. The fact that our theoretical and predictive approach correctly reproduces the experimental low-pressure results gives considerable credit to our predictions.

The first room-temperature binary multiferroic material is thus within reach: CuO at pressures of 20-40 GPa. To be practical for technical applications the high-pressure form of CuO must be made stable at ambient conditions. To achieve this, there are at least two strategies. Very special for CuO is the possibility to stabilize its high pressure form at a nanoscale level by applying high-energy ion irradiation at high pressures[31]. In such experiments, the quenched high-pressure structure remains even after the releasing of the pressure. Another strategy is provided by core-shell synthesis[32] according to which CuO nanoparticles are embedded in a shell material that has a negative thermal expansion coefficient, which then acts as an effective pressure medium for the CuO core.

**Acknowledgments:** X.R. thanks the IMN (Nantes) and the CCIPL (Centre de Calcul Intensif des Pays de la Loire) for computing facilities. S.K. Acknowledges support from DST, India.

**Author contribution:** X.R. designed and performed the DFT calculations. X.R. wrote the paper with contributions from all co-authors. S.K. performed the MC simulations. All co-authors contributed to analyzing and discussing the results.

**Supplementary Online Materials:**

Materials and Methods

Figures S1-S5

References

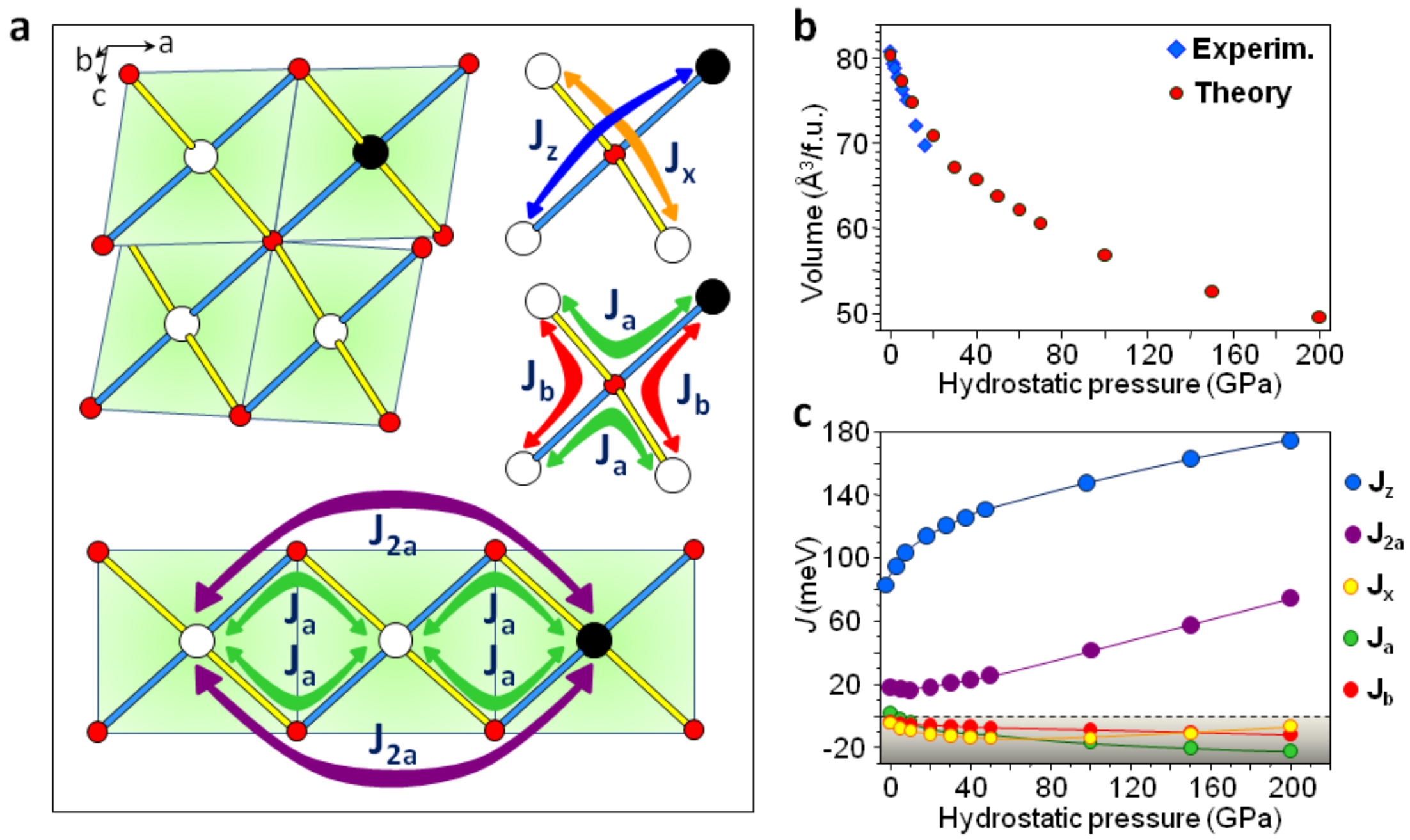

**Fig. 1**. **High-pressure evolution of the structural and magnetic properties of CuO. (a)** Schematic view of the tetrahedral environment of oxygen atoms in CuO and definition of the largest ($J_z$) and smaller magnetic super-exchange couplings ($J_x$, $J_a$ and $J_b$). The super-superexchange magnetic coupling, $J_{2a}$, corresponds to the second neighbour interaction of the edge-sharing chains, defined by the first-neighbour interaction, $J_a$. Oxygen atoms are represented by small red dots, and the $Cu^{2+}$ sites are depicted as filled and open dots, representing up-spin and down-spin, respectively. **(b)** Pressure dependence of the volume of CuO. The experimental values, deduced from a Birch-Murnaghan equation of state fitted to data of nanocrystalline CuO up to a pressure of 17 GPa[17], are compared to those calculated by DFT. **(c)** Pressure dependence of the magnetic exchange couplings of CuO. Positive and negative values represent AFM and FM interactions, respectively. The J's in the grey area are FM.

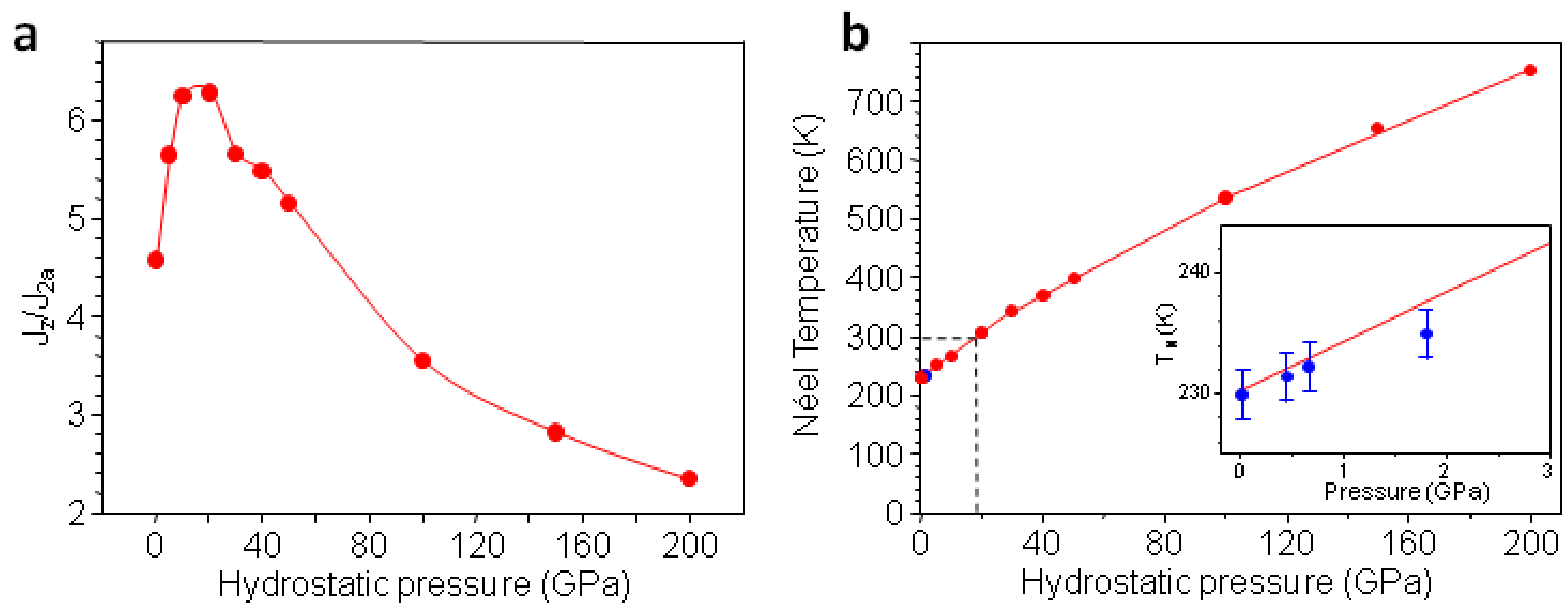

**Fig. 2. High-pressure evolution of the effective magnetic dimensionality and Néel temperature ($T_N$) of CuO. (a)** Ratio between the two largest magnetic exchange interactions ($J_z/J_{2a}$). The 1D-character of the magnetic structure is first enhanced with pressure (up to 20 GPa) and is then reduced. **(b)** Pressure dependence of the Néel temperature of CuO. Experimental data (in blue) measured up to 1.8 GPa[19] are compared to the result of the semi-empirical RPA expression (in red) for quasi-1D antiferromagnets[18].

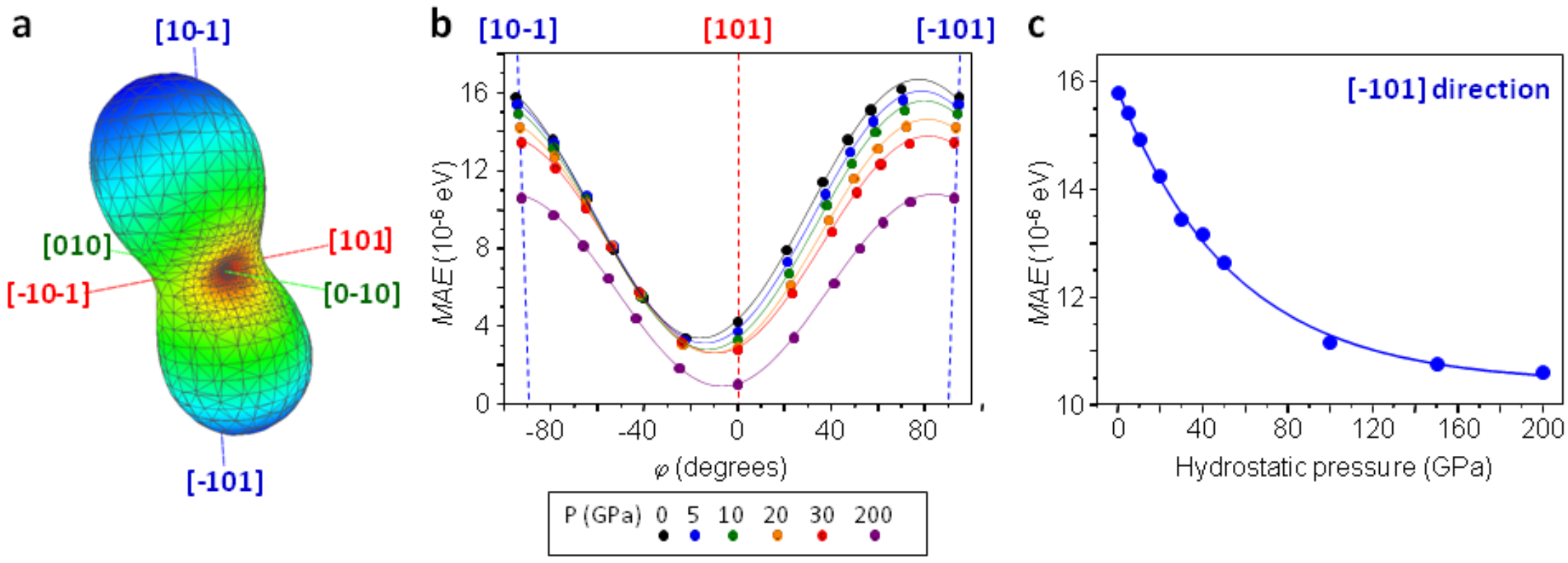

**Fig. 3. High-pressure evolution of the magneto-crystalline anisotropy of CuO. (A-C)**. Magneto-crystalline anisotropy energy (MAE) of CuO calculated for the ground-state $AF_1$ magnetic structure[7] where MAE = E[uvw] – E[010] and E[uvw] is the energy deduced from spin-orbit calculations with magnetization along the [uvw] crystallographic direction. **(a)** The 3D-shape of MAE shows that the easy axis of magnetization at 0 GPa is the b-axis, *i.e.* [010] direction of the crystallographic cell. **(b)** MAE in the plane normal to the b-axis, which is reduced by pressure. **(c)** Exponential decay of MAE with pressure, as illustrated for the [-101] direction.

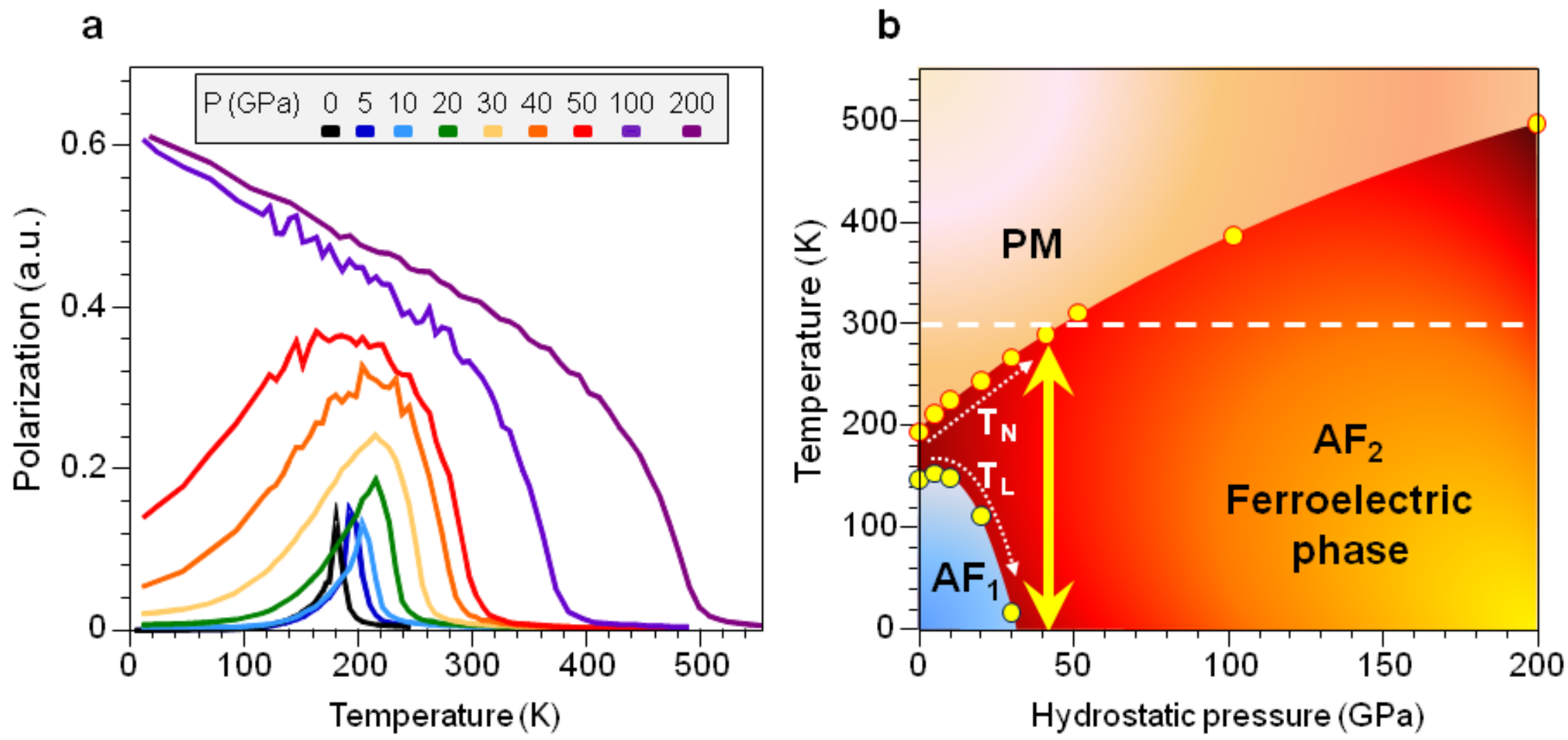

**Fig. 4. Temperature-Pressure phase diagram of the magnetic model of CuO.** Magnetic and ferroelectric properties of CuO as determined by Monte Carlo calculations, based on the microscopic magnetic interactions. **(a)** Temperature dependence of the ferroelectric polarization, which is proportional to the calculated spin current, for different values of hydrostatic pressure. **(b)** Temperature-pressure magnetic phase diagram of CuO. The room-temperature is indicated by the horizontal white dashed-line and the giant stabilization of the AF2 ferroelectric phase of CuO is highlighted by the vertical yellow double arrow.

# Supplementary Online Material for

**High-pressure cupric oxide: a room-temperature multiferroic**

Xavier Rocquefelte[1*], Karlheinz Schwarz[2], Peter Blaha[2], Sanjeev Kumar[3], Jeroen van den Brink[4]

*Correspondence to: Xavier.Rocquefelte@cnrs-imn.fr.

**This PDF file includes:**

Materials and Methods

Figs. S1 to S5

References

## Materials and Methods

### S1. DFT calculations

The Density Functional Theory (DFT) calculations have been carried out by using two different codes: Vienna Ab initio Simulation Package (VASP)[1] for the geometry optimization at the different pressure values and WIEN2k program package[2] for the calculation of the magnetic exchange, $J_{ij}$, and magnetocrystalline anisotropy energy, MAE, values.

*DFT geometry optimizations:*

A 16 formula units cell has been used, i.e. 2a×2b×2c, with a, b and c the crystallographic cell parameters. The ground state magnetic order ($AF_1$) has been considered for the geometry optimization. The parameters used in the VASP calculations are the following. We have used the GGA+U approach with $U_{eff}$ = 6.5 eV for the Cu(3d) states, as in our previous investigation[3,4]. It allows having a proper description of the structural properties of CuO. The wave functions are expanded in a plane wave basis set with kinetic energy below 500 eV. The VASP package is used with the projector augmented wave (PAW) method of Blöchl[5]. The integration in the Brillouin Zone is done by the Methfessel-Paxton method[6] on a 3×3×3 set of k-points determined by the Monkhorst-Pack scheme[7]. All atoms were then allowed to relax by following a conjugate gradient minimization of the total energy scheme ($3\times10^{-2}$ eV/ Å).

*DFT estimation of the $J_{ij}$ values:*

Based on the optimized atomic structures and for each pressure (from 0 to 200 GPa), DFT calculations were performed using the WIEN2k program package with the PBE0[8] hybrid functional as previously discussed in ref. 9 for 8 and 32 f.u. cells. The $J_{ij}$ values have been deduced from a least-squares fit procedure and the quality of the fits is shown in Figs. S1 and S2.

*DFT estimation of the magnetic anisotropy:*

The magnetic anisotropy energy (MAE) has been estimated for the $AF_1$ ground-state magnetic structure, using the code WIEN2k with the PBE0 hybrid functional and including the spin-orbit coupling. MAE corresponds to an energy difference between two directions of the magnetization density. Here we use the [010] direction, i.e. the easy axis, as the reference:

MAE = E[uvw] – E[010]

E[uvw] is the energy deduced from spin-orbit calculations with magnetization along the [uvw] crystallographic direction. It should be noticed that MAE is very sensitive to the k-mesh. The quality of the k-mesh has been carefully chosen, leading to the use of a 5×12×6 set of k-points for the 8 f.u. cell.

**S2. Estimation of $T_N$ based on the RPA formula:**

$$J' = T_N \Big/ \left[ 4c \sqrt{ ln\left(\frac{\lambda J}{T_N}\right) + \frac{1}{2} ln ln\left(\frac{\lambda J}{T_N}\right) } \right]$$

The above equation has been developed for the estimation of $T_N$ of quasi-1D antiferromagnetic Heisenberg model on a cubic lattice with *J* and *J'*, the intrachain and interchain couplings, respectively[10]. The related ground state (GS) magnetic order leads to the following energy expression:

E(GS) = $J + 2J'$

Although cupric oxide is a quasi-1D magnetic system, it exhibits a more complex magnetic order due to the low-symmetry of its atomic structure (monoclinic space group: C2/c). As a consequence, its ground state magnetic order ($AF_1$) leads to the following energy expression:

$E(AF_1) = J_z - J_x + J_2$ with $J_2 = J_{2a} + J_{2b} + J_{2c}$

with $J_{2a}$ the predominant super-superexchange interaction. For more details, see the refs. 4 and 9. Considering E(GS) = $E(AF_1)$ and $J_z$ as the intrachain coupling, i.e. $J_z = J$, we can define J' as:

$2J' = - J_x + J_2$

Our detailed results are illustrated in Fig. S3.

## S3. Classical Monte Carlo simulations:

We use the Metropolis algorithm to perform Markov Chain Monte Carlo simulations on classical spins. The simulations are started with a completely random spin configuration at high temperature. Due to the presence of many competing interactions and nearly degenerate ground states, the simulations require a large number of equilibration and averaging steps. We use ~$10^6$ Monte Carlo steps for equilibration and a similar number of steps for averaging at each temperature. The temperature is then reduced in small steps (~5K) and the system is allowed to anneal towards the ground-state spin configuration. The simulations are carried out on lattices with $N = 12^3$ sites. We have checked the stability of our results for larger sizes (up to $N=32^3$) for selected values of pressure (see Fig. S5).

The model Hamiltonian used in this work is similar to the one used in ref. 3. In addition we introduce a phenomenological multiaxial anisotropy term of the form[11]:

$$H_{MA} = B\sum_i \left(S_i^y S_i^z\right)^2$$

The uniaxial anisotropy parameter $\lambda = 0.02$ is kept constant. The multiaxial anisotropy B decreases exponentially with increasing pressure; we use $B = 500\, e^{-P/10}$ (B = 500 for P = 0, and B = 24.9 for P = 30). The large value of B at P = 0 is required to obtain the narrow range of stability of $AF_2$ state at high temperatures. Although uniaxial anisotropy term is also decreasing with pressure, this does not lead to any crucial changes in the phase diagram shown in Fig. 4. We also keep the DM coupling fixed to D = 0.8Dc. The procedure used to estimate $T_L$ and $T_N$ for 2 pressures (0 and 30 GPa) is shown in Fig. S4.

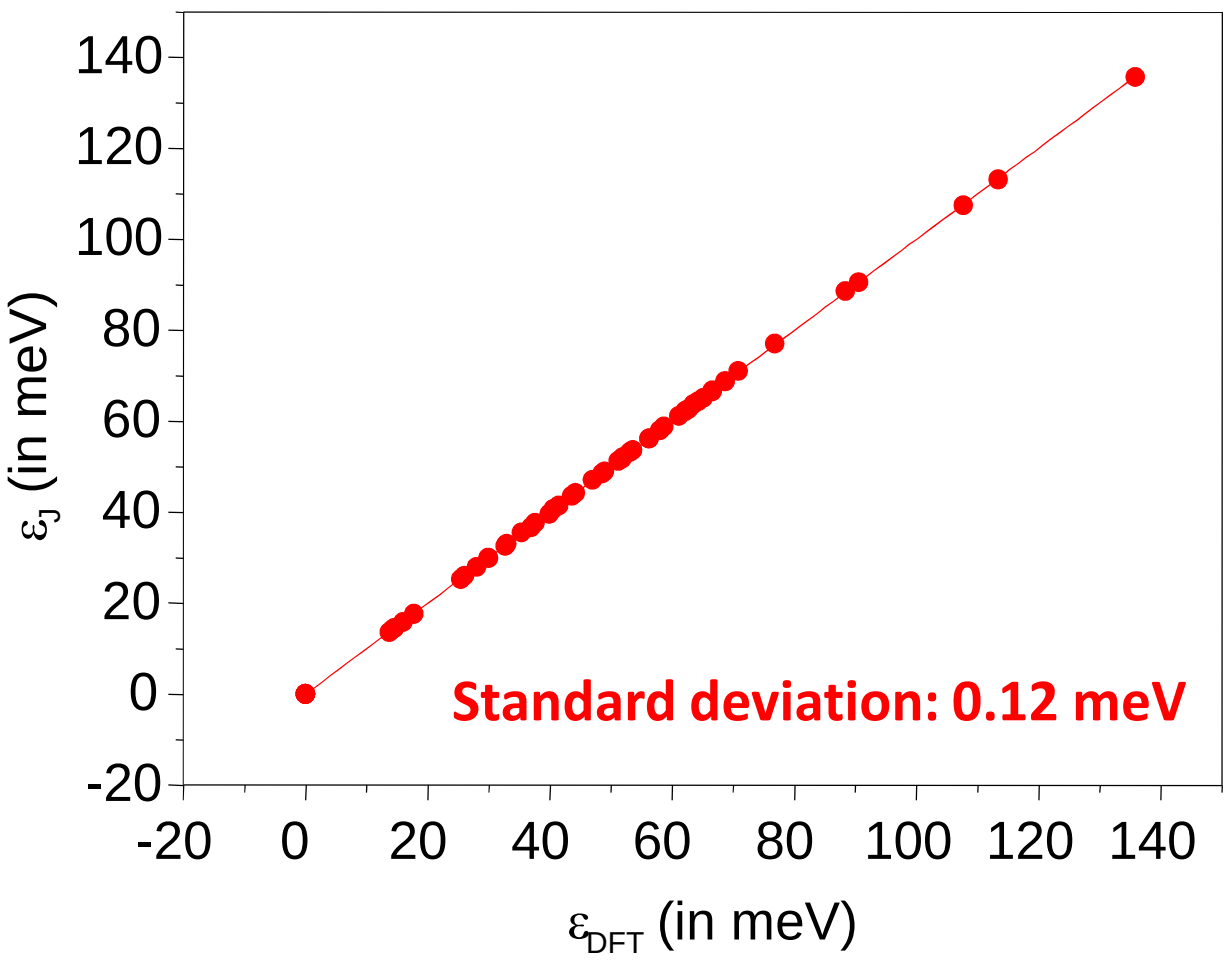

**Fig. S1.** Graphical representation of the quality of the least-squares fit procedure for the 8 formula units models. $\varepsilon_{DFT}$ and $\varepsilon_J$ are, respectively, the relative energies (with respect to AF1) deduced from the DFT calculations and the J parameters.

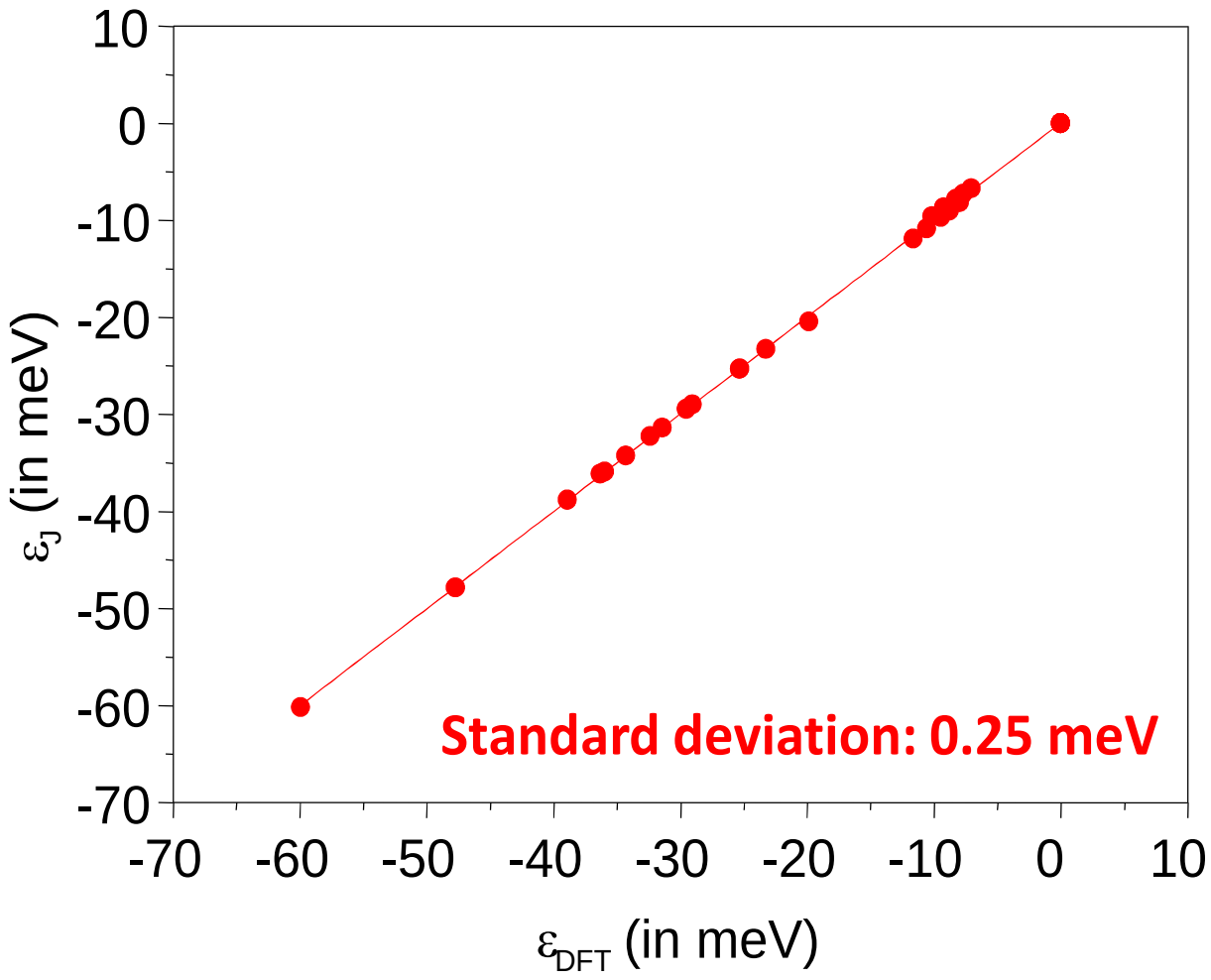

**Fig. S2.** Graphical representation of the quality of the least-squares fit procedure for the 32 formula units models. $\varepsilon_{DFT}$ and $\varepsilon_J$ are, respectively, the relative energies (with respect to AF1) deduced from the DFT calculations and the J parameters.

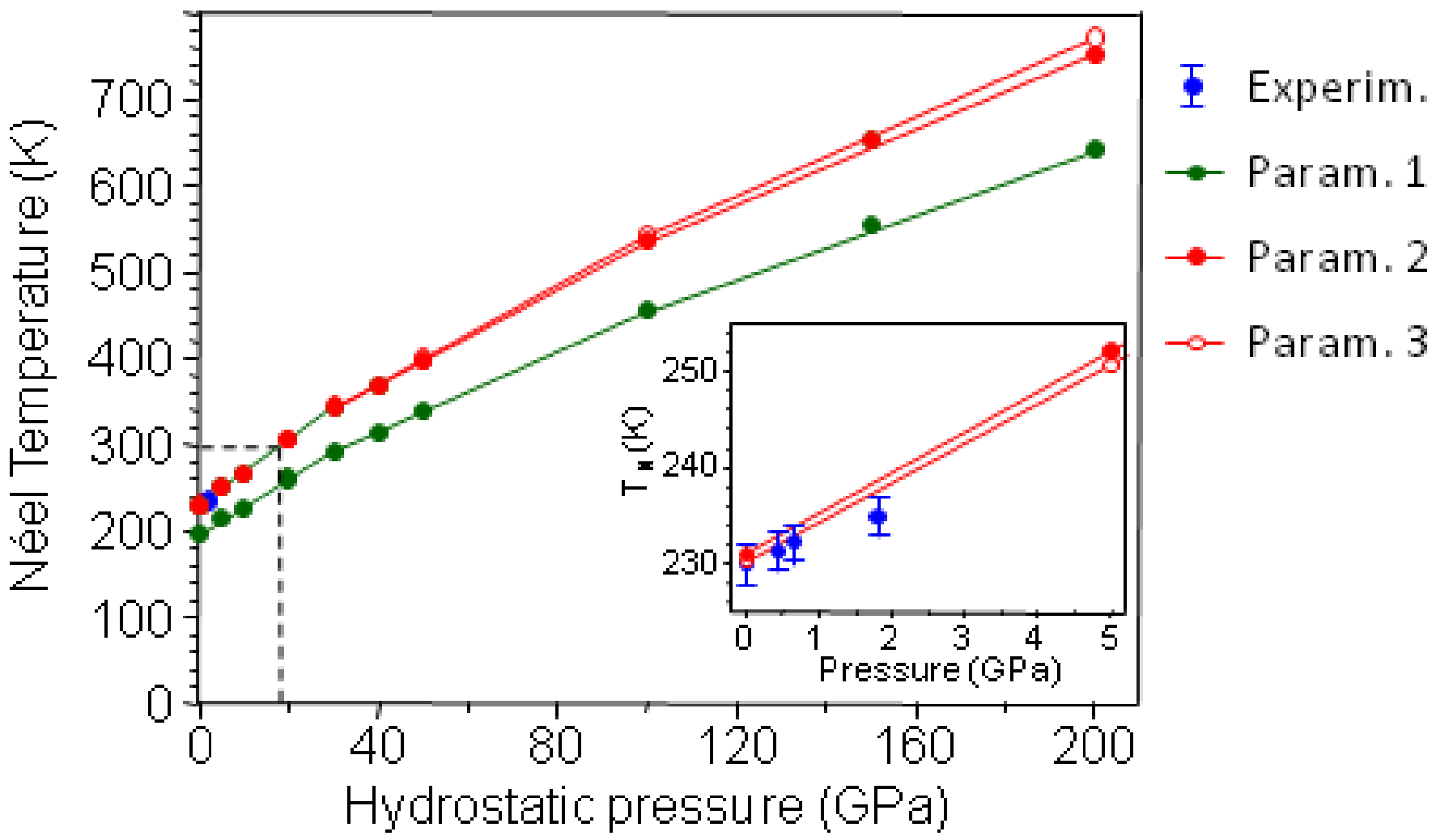

**Fig. S3. High-pressure evolution of the Néel temperature ($T_N$) of CuO.** Pressure dependence of the Néel temperature of CuO. Experimental data measured up to 1.8 GPa are compared to theoretical ones deduced from an analytical expression developed for S = 1/2 quasi-1D Heisenberg antiferromagnets[10]. Three parameterizations are used, the original one with c = 0.233 and λ = 2.6 (param.1) and two modified forms, with c = 0.284 and λ = 2.6 (param.2) and with c = 0.233 and λ = 8.4 (param.3). The black dash-line evidences that $T_N$ is reaching room-temperature at about 20 GPa (from both param.2 and 3).

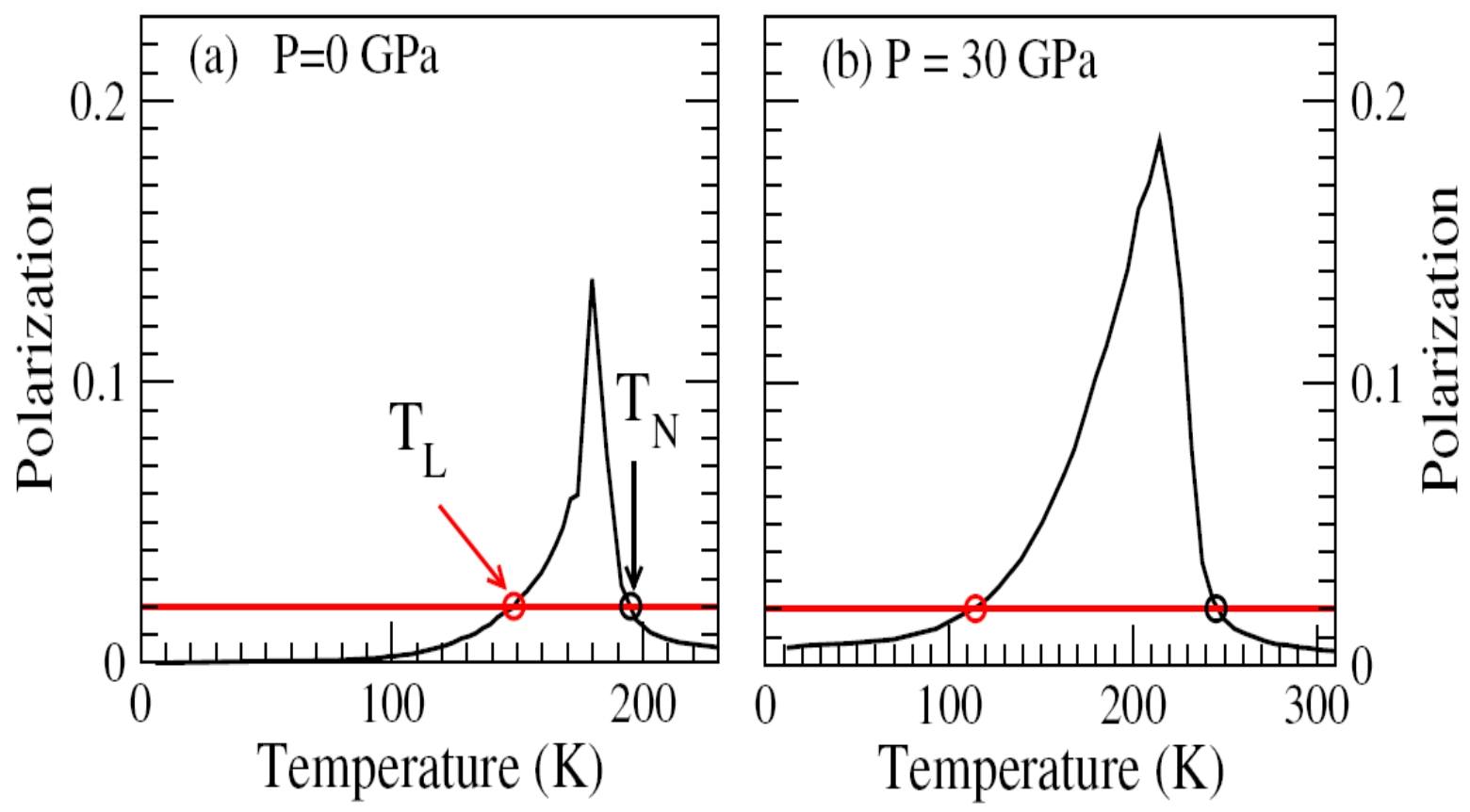

**Fig. S4.** Temperature dependence of the spin current (which is proportional to ferroelectric polarization) for two sets of parameters corresponding to (a) P = 0 GPa, and (b) P = 30 GPa. An increase in spin current upon decreasing temperature is an indication of the onset of a non-collinear ferroelectric phase. The decrease of spin-current below a cutoff value is defined at the transition to a collinear state. We show the red horizontal line (P = 0.02) as the cutoff value used for inferring $T_N$ and $T_L$.

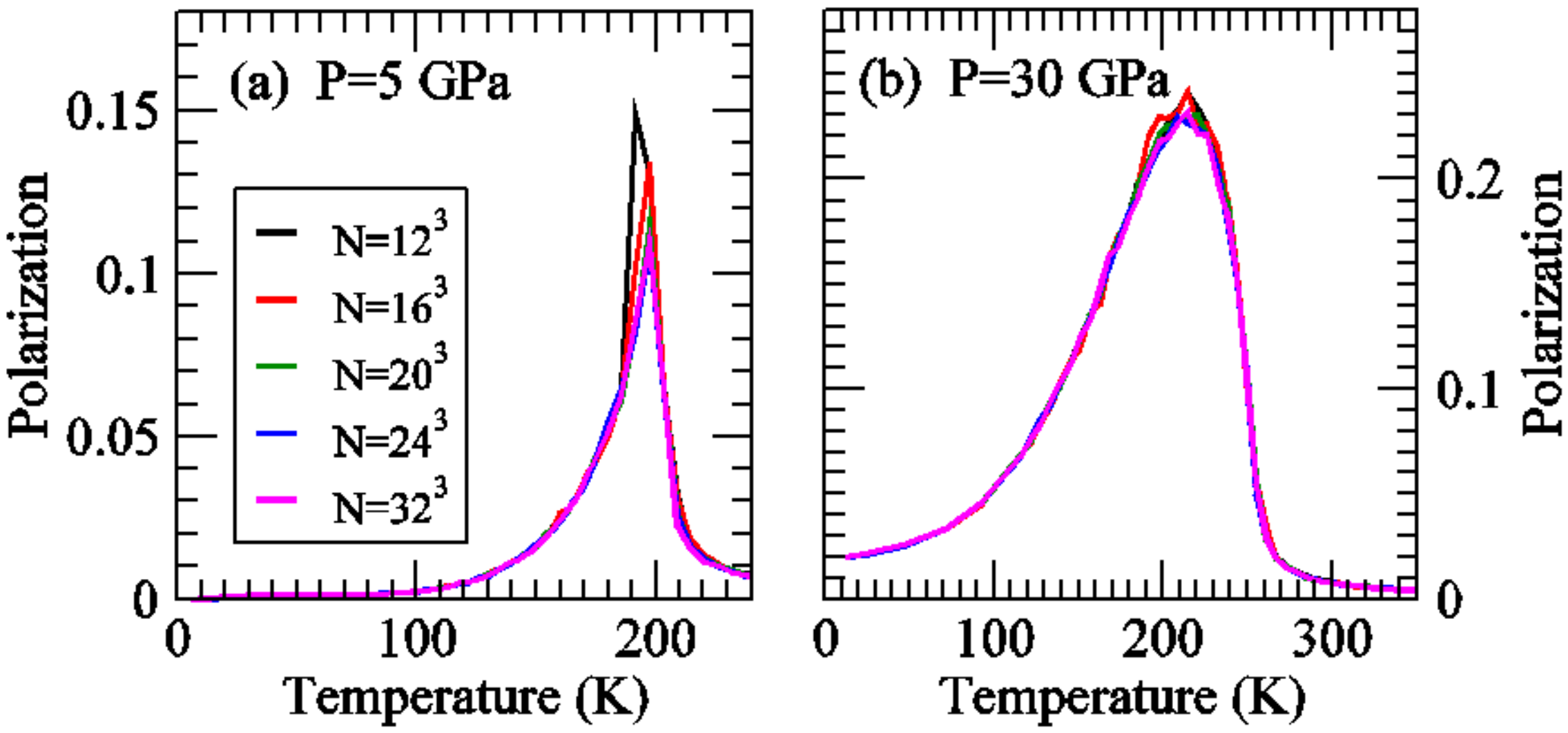

**Fig. S5.** Stability of our MC simulations as a function of the lattice size (up to N=$32^3$) for selected values of pressure, i.e. 5 and 30 GPa.